# Aerocapture Design Reference Missions for Solar System Exploration: from Venus to Neptune

Athul Pradeepkumar Girija [1][**][**]

[1]School of Aeronautics and Astronautics, Purdue University, West Lafayette, IN 47907, USA

**ABSTRACT**

Aerocapture is the technique of using planetary atmospheres to decelerate a spacecraft in a single pass to achieve nearly fuel-free orbit insertion. Aerocapture has been extensively studied since the 1980s but has never been flown yet. The entry conditions encountered during aerocapture are strongly destination dependent, and performance benefit offered by aerocapture is also destination dependent. Aerocapture is applicable to all atmosphere-bearing destinations with the exception of Jupiter and Saturn, whose extreme entry conditions make aerocapture infeasible. A recent study by the NASA Science Mission Directorate highlighted the need for baseline design reference missions, as a starting point for system level architecture studies. The present study uses the Aerocapture Mission Analysis Tool (AMAT) to compile a list of design reference missions at Venus, Earth, Mars, Titan, Uranus, and Neptune. These reference missions can provide an initial assessment of the feasibility of aerocapture for a proposed mission, and provide intial baseline values for more detailed system studies. The reference mission set provides a quick estimate of the entry conditions, control requirements, and aero-thermal loads for architectural level studies.

[****] To whom correspondence should be addressed, E-mail: athulpg007@gmail.com

## I. INTRODUCTION

Aerocapture is the technique of using planetary atmospheres to decelerate a spacecraft in a single pass to achieve nearly fuel-free orbit insertion. Compared to conventional propulsive insertion, aerocapture can significantly reduce the required propellant mass required, allowing for smaller and less expensive spacecraft for planetary missions [1]. For the inner planets, aerocapture can enable insertion of small low-cost satellites into low circular orbits around Mars and Venus [2]. For the outer planets, aerocapture can greatly increase the delivered mass to orbit and enable significantly shorter time of flight missions than propulsive insertion [3]. Aerocapture is applicable to all atmosphere-bearing destinations except Jupiter and Saturn, whose extreme entry conditions make aerocapture infeasible [4]. Figure 1 shows an illustration of the aerocapture maneuver. Aerocapture has been extensively studied since the 1980s but has never been flown yet. The entry conditions encountered during aerocapture are strongly destination dependent, and performance benefit offered by aerocapture is also destination dependent [5, 6]. A recent study by the NASA Science Mission Directorate highlighted the need for design reference missions, as a starting point for system level architecture studies [7]. The present study uses the Aerocapture Mission Analysis Tool (AMAT) to compile a list of design reference missions at Venus, Earth, Mars, Titan, Uranus, and Neptune [8, 9]. The reference mission set provides a quick estimate of the entry conditions, control requirements, and aero-thermal loads for architectural level studies.

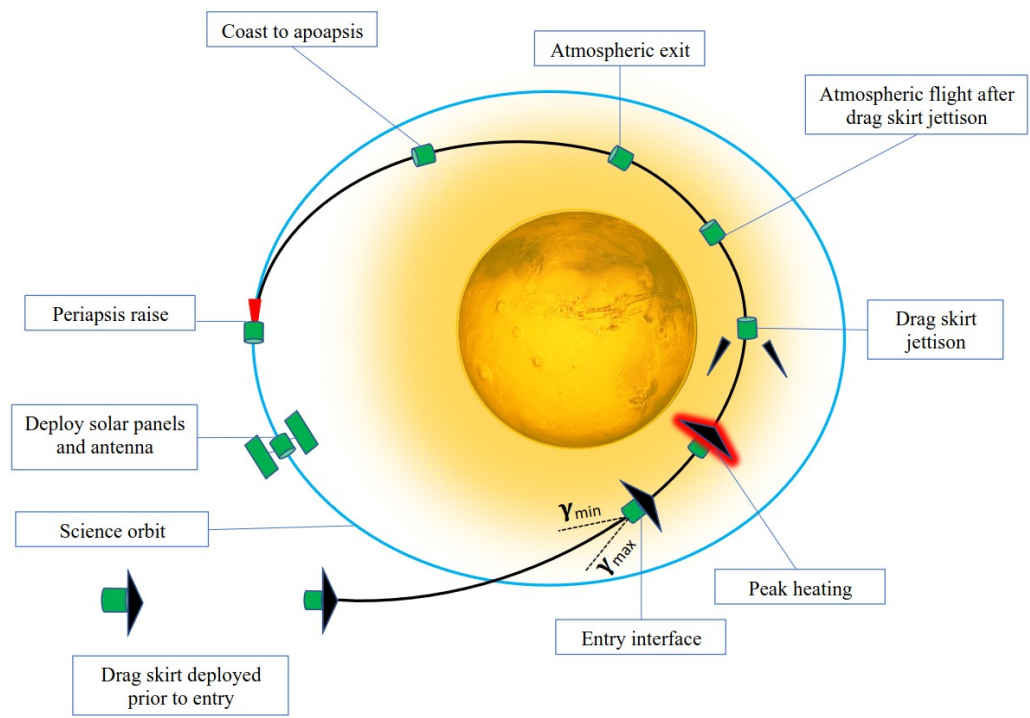

Figure 1. Schematic illustration of the aerocapture maneuver.

## II. VENUS

Aerocapture at Venus has been studied using both rigid aeroshells (eg: Apollo) as well as low-ballistic coefficient deployable systems. Due to the demanding aero-thermal conditions posed by the thick Venusian atmosphere, rigid aeroshells require significant thermal protection system (TPS) mass and is not attractive for aerocapture [10]. Aerobraking instead is the preferred option, and is baselined for all planned missions. However, using drag modulation aerocapture to insert small satellites into Venus orbit is very attractive and has been studied extensively [11, 12]. The low-ballistic coefficient entry system keeps the heating rates low [13], and has applications for delivering small independent secondary payloads to Venus orbit, small payloads as components of Flagship missions [14], and sample return missions from the clouds [15]. The proposed baseline design reference mission for Venus aerocapture uses a 1.5 m diameter vehicle with mass=53 kg, $\beta_1$=20 kg/m$^2$, ballistic coefficient ratio $\beta_2/\beta_1$=7.5 to deliver a 25 kg orbiter to a 200 x 2000 km orbit. The entry speed is 10.8 km/s and the aerocapture corridor is [-5.53, -5.11] deg., with a width of 0.42 deg. Figure 2 shows the Venus design reference trajectory. The peak deceleration is about 7g, and the peak stagnation point-heat rate is about 150 W/cm$^2$, which is within the tested limits of the ADEPT carbon cloth TPS. The $\Delta V$ offered by the maneuver is 3113 m/s, and the periapsis raise $\Delta V$ is about 30 m/s.

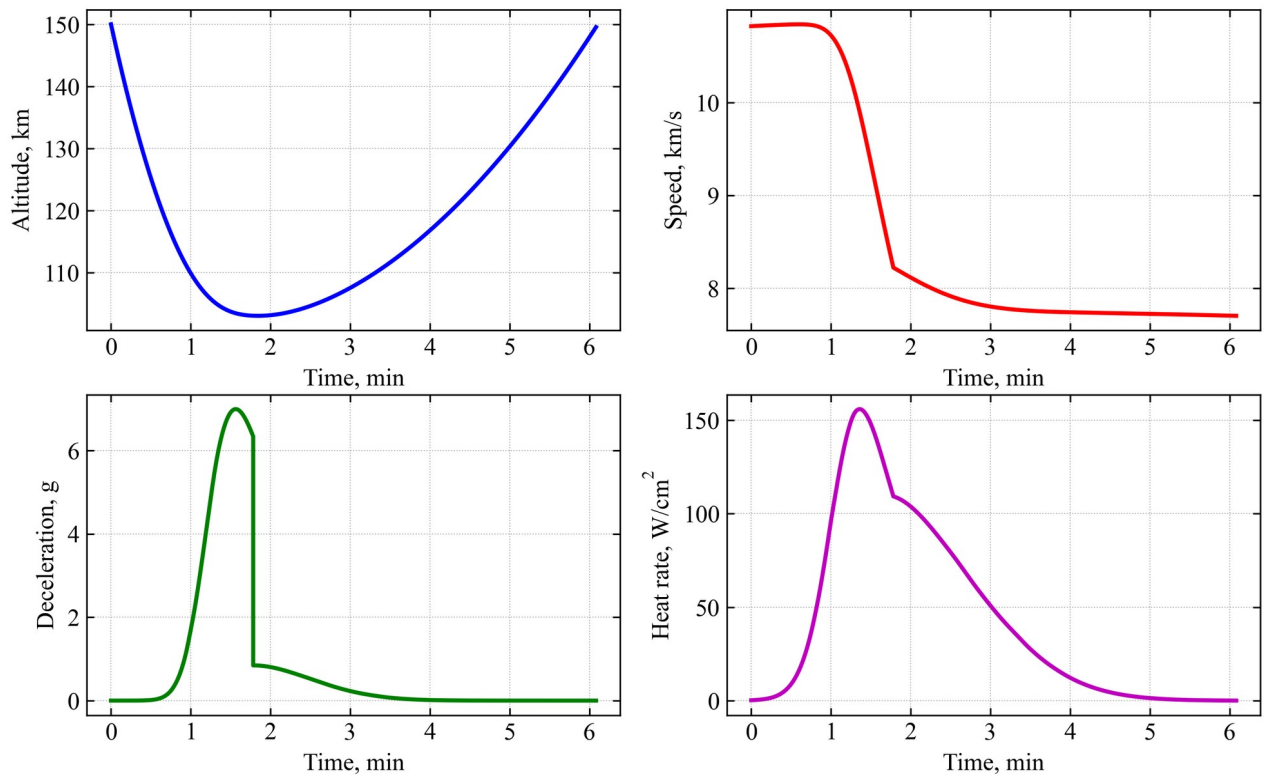

Figure 2. Venus aerocapture design reference mission trajectory.

## III. EARTH

Aerocapture at Earth has been extensively studied since the 1980s for various technology demonstration missions such as the AFE, ST-7, and ST-9 flight opportunities, though none of them were eventually flown [16]. There is considerable renewed interest in the last few years for its applications to the Aerocapture Flight Test Experiment (AFTE) which was first proposed by Werner and Braun [17], and has since then been refined considerably by NASA researchers towards a low-cost technology demonstration flight mission [18]. The small spacecraft (under 200 kg) would launch as a secondary payload on a GTO mission, perform a burn to enter the atmosphere, and use the atmospheric pass to change its orbit from GTO to a smaller elliptic orbit. The mission would demonstrate all aspects of drag modulation aerocapture, bring the technology to flight readiness with immediate applications to Mars and Venus missions, and potentially pave the way for more ambitious outer planet missions with aerocapture. The proposed reference design uses the same vehicle design as for Venus and targets a 2000 km apoapsis altitude. The entry speed is 10.6 km/s and the aerocapture corridor is [-4.87, -4.18] deg., with a width of 0.69 deg. Figure 3 shows the Earth aerocapture design reference mission trajectory. The peak deceleration is about 5g, and the peak stagnation point-heat rate is about 125 W/cm$^2$. The ΔV offered by the maneuver is 2300 m/s, and the periapsis raise ΔV is about 36 m/s.

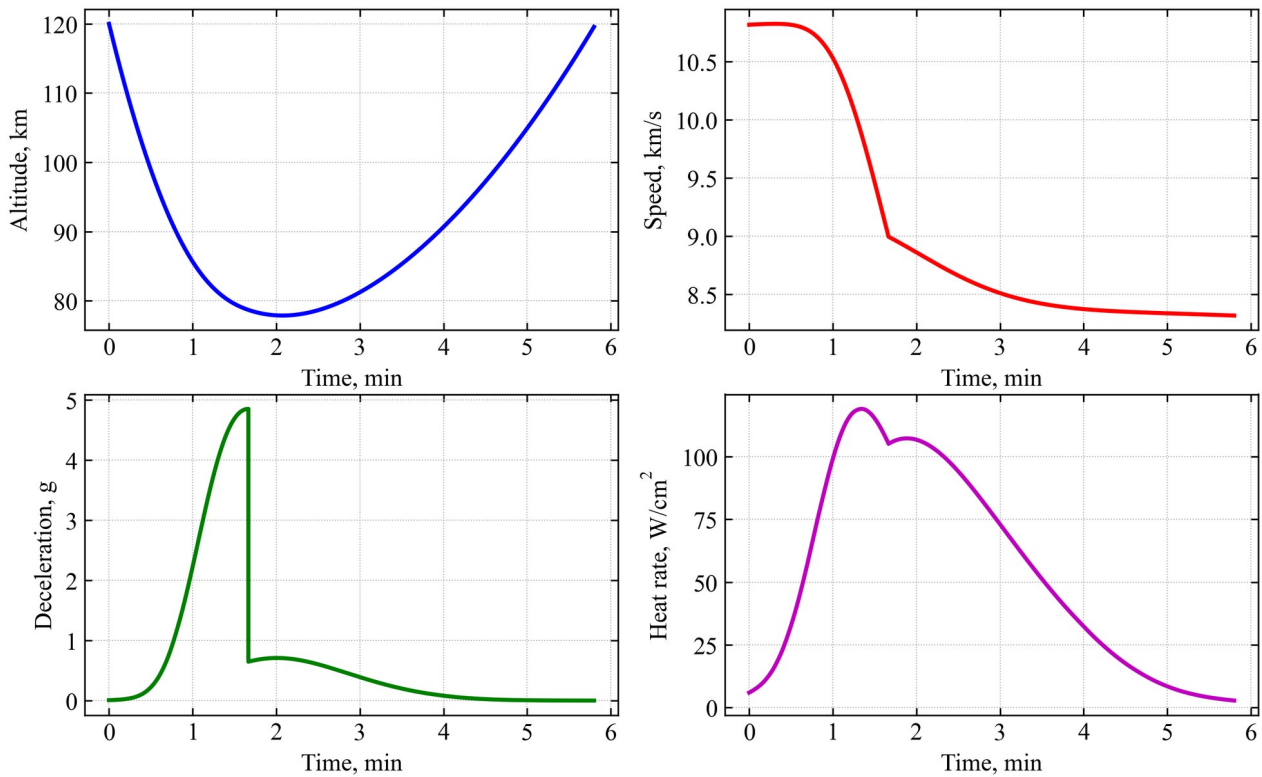

Figure 3. Earth aerocapture design reference mission trajectory.

## IV. MARS

Aerocapture at Mars has also been studied since the 1980s particularly with its application for Mars Sample Return (MSR) orbiters. The thin Martian atmosphere presents a relatively benign aero-thermal environment for aerocapture, making it the most studied destination for aerocapture. More recently, drag modulation aerocapture has received renewed interest for insertion of small satellites around Mars [19]. The low gravity of Mars combined and the extended atmosphere makes it more attractive for aerocapture in terms of larger corridor width and lower heating rates compared to Earth and Venus. The low heating rates at Mars and the frequent launch opportunities make Mars an ideal candidate for a low-cost aerocapture technology demonstration mission outside of the Earth [20]. The reference design uses the same vehicle as for Venus and targets a 2000 km apoapsis altitude. The entry speed is 5.4 km/s and the aerocapture corridor is [-9.86, -8.78] deg., with a width of 1.09 deg. Note the trend in corridor widths across Venus (0.42 deg.), Earth (0.69 deg.) and Mars (1.09 deg.) for the same vehicle design and the target orbit, illustrating the effect of differences in the atmospheric structure on the aerocapture corridor width. Figure 4 shows the Mars aerocapture design reference mission trajectory. The peak deceleration is about 2.5g, and the peak stagnation point-heat rate is about 20 W/cm$^2$. The ΔV offered by the maneuver is 1760 m/s, and the periapsis raise ΔV is about 33 m/s.

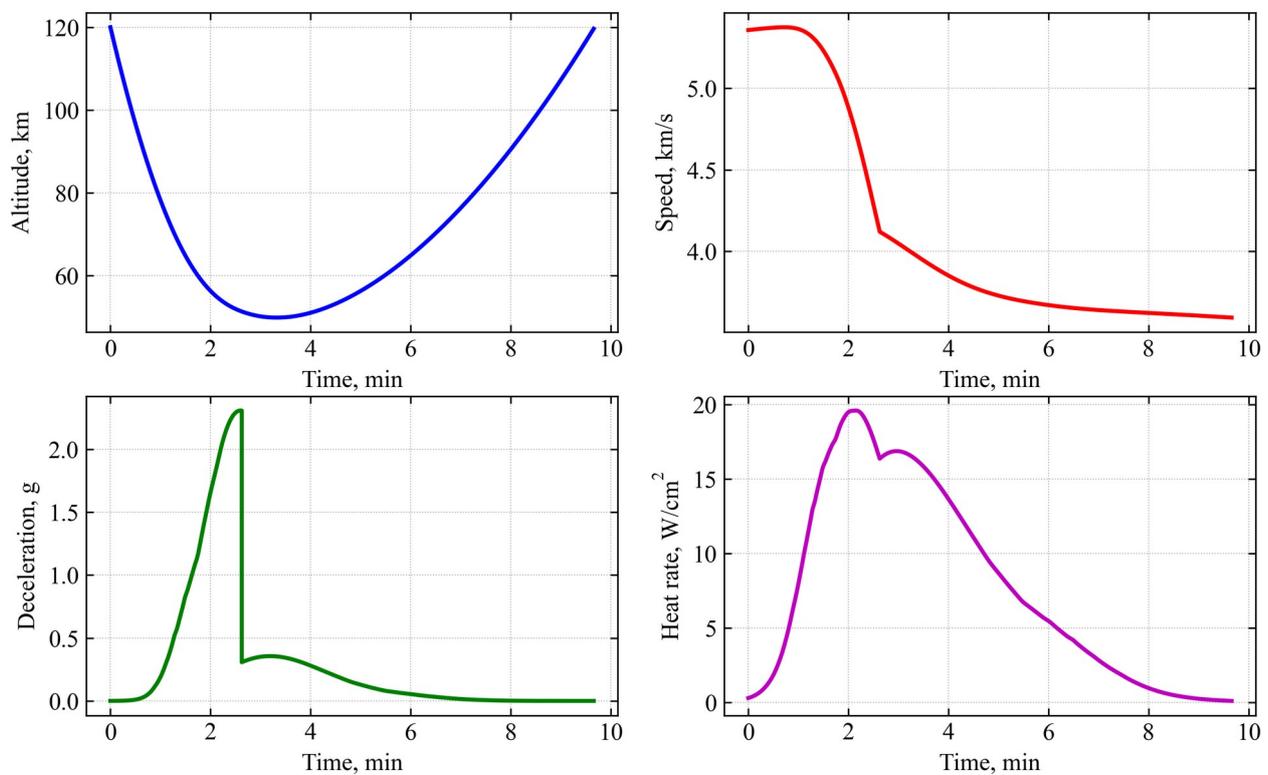

Figure 4. Mars aerocapture design reference mission trajectory.

## V. TITAN

With its low-gravity and greatly extended atmosphere, Titan presents the most attractive destination for aerocapture in terms of the largest corridor width and the lowest heating rate. The benign environment makes both rigid aeroshells and deployable systems viable for aerocapture at Titan. In the early 2000s, using blunt-body aeroshells was extensively studied as part of the Aerocapture Systems Analysis Study [21]. More recently, attention has focused on drag modulation at Titan due to its simplicity and lower cost compared to rigid aeroshells. The proposed baseline design reference mission for Titan drag modulation aerocapture uses a 12 m diameter vehicle with mass=5700 kg, $\beta_1$=30 kg/m$^2$, ballistic coefficient ratio $\beta_2/\beta_1$=4.14 to deliver a 2600 kg orbiter to a 1700 km circular orbit. The entry speed is 7.3 km/s and the aerocapture corridor is [-37.3, -34.4] deg., with a width of 1.89 deg. Figure 5 shows the Titan aerocapture design reference mission trajectory. Note the considerably longer duration of the aerocapture maneuver at Titan compared to the inner planets, and the considerably wider aerocapture corridor. The peak deceleration is about 3.5g, and the peak stagnation point-heat rate is about 30 W/cm$^2$. The $\Delta$V offered by the maneuver is 5750 m/s, and the periapsis raise $\Delta$V is about 150 m/s. With the need for an orbiter around Titan following the Dragonfly mission, ADEPT drag modulation aerocapture has applications for a Titan orbiter which fits within New Frontiers [22].

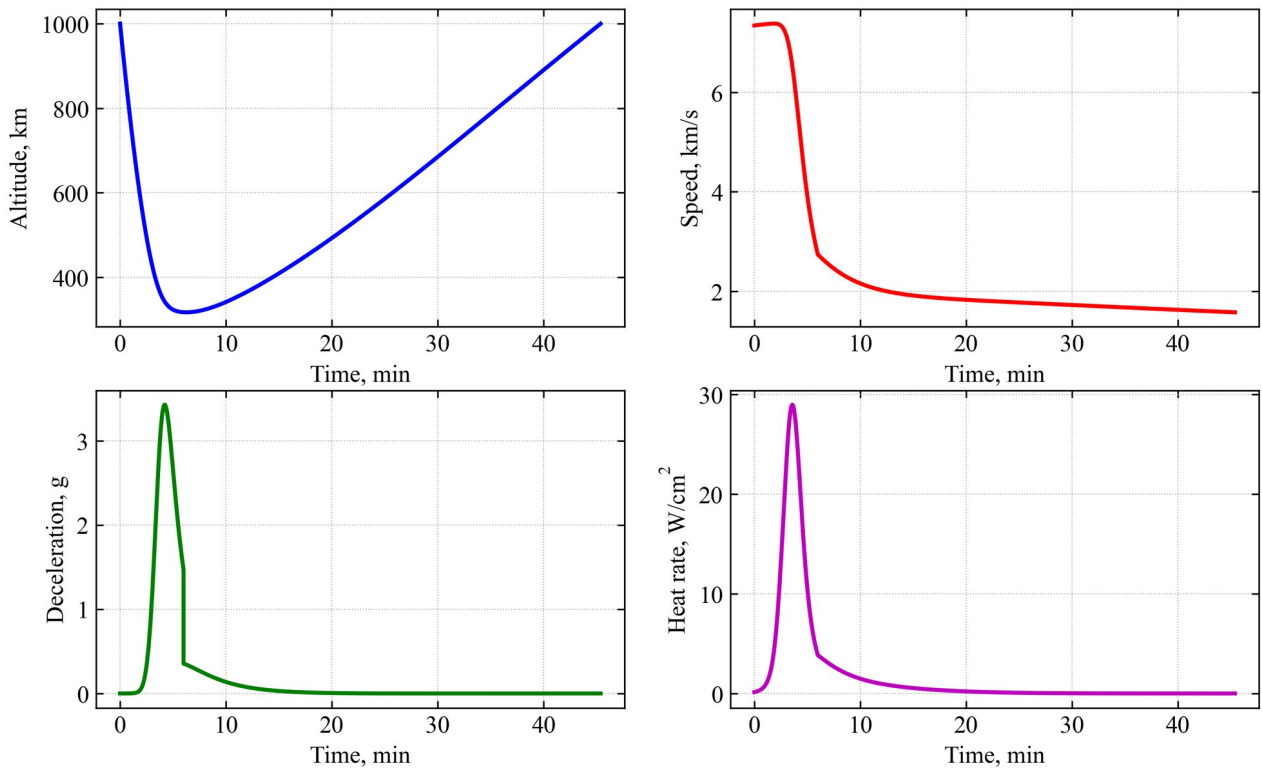

Figure 5. Titan aerocapture design reference mission trajectory.

## VI. URANUS

Uranus' large heliocentric distance (19 AU) presents significant mission design challenges for orbit insertion. For cost and risk reasons, current baseline Uranus mission architectures do not use aerocapture [23, 24]. However, aerocapture can significantly increase the delivered mass to orbit and considerably reduce the flight time [25, 26]. The considerably large navigation and atmospheric uncertainties compared to the inner planets and Titan make lift modulation the preferred control method. The proposed baseline design reference mission for Uranus aerocapture uses an MSL derived aeroshell with mass=3200 kg, β=146 kg/m$^2$, L/D=0.24 to deliver a 1400 kg orbiter + 300 kg atmospheric probe to a 4000 x 500,000 km orbit. A high entry speed is chosen to maximize the available control authority with the low-L/D aeroshell. The entry speed is 29 km/s and the corridor is [-12.0, -11.0] deg., with a width of 1.00 deg. Figure 6 shows the Uranus aerocapture design reference mission overshoot and undershoot bounding trajectories. The peak deceleration is in the range of 4–10g, and the peak stagnation point-heat rate is about 1400–2000 W/cm$^2$. The ΔV offered by the maneuver is 8900 m/s, and the periapsis raise ΔV is about 80 m/s. Compared to probes which enter steep, aerocapture requires shallow entry which results in large heat loads [27]. For the reference Uranus mission, the total heat load is 230 kJ/cm$^2$, which is expected to have a TPS% of 20–25% with HEEET [28].

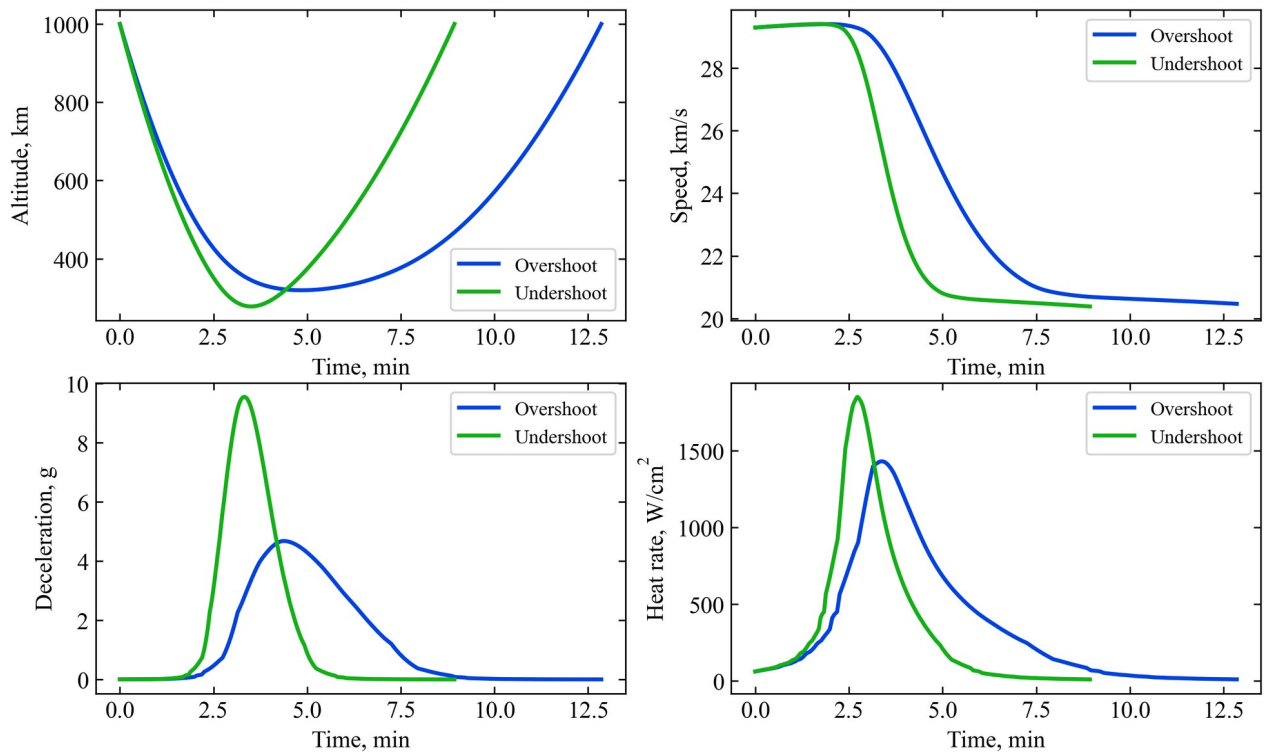

Figure 6. Uranus aerocapture design reference mission overshoot and undershoot trajectories.

## VII. NEPTUNE

At a heliocentric distance of 30 AU, the ice giant Neptune presents an even greater challenge for orbiter spacecraft than Uranus. Aerocapture at Neptune was extensively studied in the early 2000s using a mid-L/D vehicle to compensate for the large navigation and atmospheric uncertainties. However, since then it has become clear such a vehicle would not be available and attention has turned to using innovative techniques to leverage low-L/D aeroshells [29, 30, 31]. The proposed baseline design reference mission for Neptune aerocapture uses an MSL derived aeroshell with mass=3200 kg, β=146 kg/m$^2$, L/D=0.24 to deliver a 1400 kg orbiter + 300 kg atmospheric probe to a 4000 x 500,000 km orbit. As with Uranus, a high entry speed is chosen to maximize the control authority. The entry speed is 30 km/s and the aerocapture corridor is [-12.69, -11.88] deg., with a width of 0.80 deg. Figure 6 shows the Neptune aerocapture design reference mission overshoot and undershoot bounding trajectories. The peak deceleration is in the range of 4–10g, and the peak stagnation point-heat rate is about 1700–2300 W/cm$^2$, which is still within the capability of the HEEET TPS. The ΔV offered by the maneuver is 8080 m/s, and the periapsis raise ΔV is about 130 m/s.

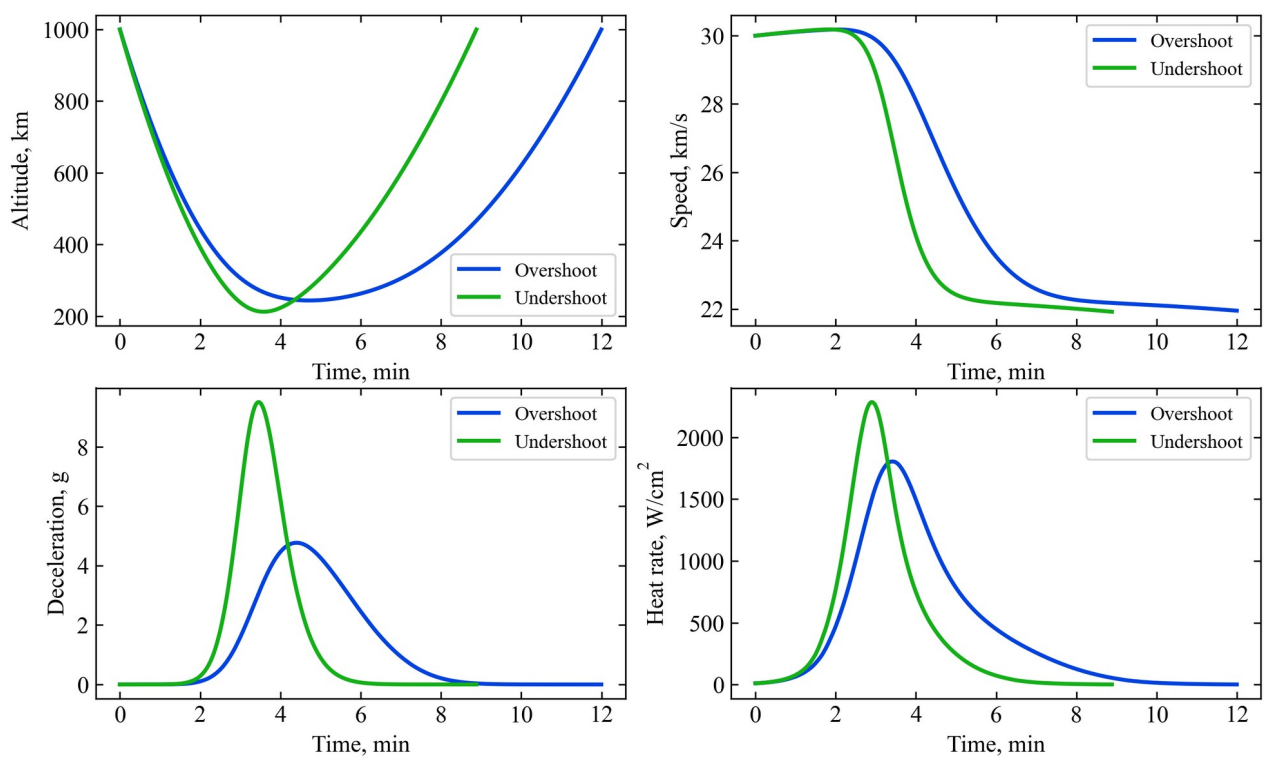

Figure 7. Neptune aerocapture design reference mission overshoot and undershoot trajectories.

## VIII. SUMMARY

Table 1 summarizes the vehicle design parameters and the entry corridor for the six design reference aerocapture missions. Table 2 summarizes the delivered mass to orbit, aero-thermal conditions, and periapse raise ΔV. These results illustrate the fact that Mars and Titan have the largest aerocapture corridors in the Solar System, and also have the lowest heating rates. Outer planet missions which have very large ΔV requirements stand to benefit the most out of aerocapture when compared to propulsive insertion which imposes severe limitations on the mission design [32].

Table 1. Comparison of vehicle design parameters and entry corridor for the reference missions.

| Mission | Control method | Entry mass, kg | $\beta_1$, kg/m$^2$ | $\beta_2/\beta_1$ or L/D | Entry speed, km/s | Corridor, deg. | Width, deg |
|---|---|---|---|---|---|---|---|
| Venus Smallsat | Drag modulation | 53 | 20 | 7.5 | 10.8 | [-5.53, -5.11] | 0.42 |
| Earth Smallsat | Drag modulation | 53 | 20 | 7.5 | 10.6 | [-4.87, -4.18] | 0.69 |
| Mars Smallsat | Drag modulation | 53 | 20 | 7.5 | 5.4 | [-9.86, -8.78] | 1.09 |
| Titan NF | Drag modulation | 5700 | 30 | 4.14 | 7.3 | [-37.3, -34.4] | 1.89 |
| Uranus Flagship | Lift modulation | 3200 | 146 | 0.24 | 29.0 | [-12.0, -11.0] | 1.00 |
| Neptune Flagship | Lift modulation | 3200 | 146 | 0.24 | 30.0 | [-12.7, -11.9] | 0.80 |

Table 2. Comparison of mass delivered to orbit, aero-thermal conditions, and periapse raise ΔV.

| Mission | Orbit size, km | Mass to orbit, kg | Aerocapture ΔV, m/s | g-load | Heat rate, W/cm$^2$ | Periapse raise ΔV, m/s |
|---|---|---|---|---|---|---|
| Venus Smallsat | 200 x 2000 | 25 | 3113 | 7.0 | 150 | 30 |
| Earth Smallsat | 200 x 2000 | 25 | 2300 | 5.0 | 125 | 36 |
| Mars Smallsat | 200 x 2000 | 25 | 1760 | 2.5 | 20 | 33 |
| Titan NF | 200 x 2000 | 2600 | 5750 | 3.5 | 30 | 150 |
| Uranus Flagship | 4000 x 500,000 | 1700 | 8900 | 4–10 | 1400–2000 | 80 |
| Neptune Flagship | 4000 x 500,000 | 1700 | 8080 | 4–10 | 1700–2300 | 130 |

## IX. CONCLUSIONS

The study compiled a list of design reference missions for aerocapture at Venus, Earth, Mars, Titan, Uranus, and Neptune. These reference missions can provide an initial assessment of the feasibility of aerocapture for a proposed mission, and provide intial baseline values for more detailed system studies, The reference mission set provides a quick estimate of the entry conditions, control requirements, and aero-thermal loads for architectural level studies.

## DATA AVAILABILITY

The results presented in the paper can be reproduced using the open-source Aerocapture Mission Analysis Tool (AMAT) v2.2.22. The data and code used to make the study results will be made available by the author upon request.